\documentclass[fleqn,10pt]{wlscirep}
\title{Chaotic, informational and synchronous behaviour of multiplex networks}

\author[1,*,+]{M. S. Baptista}
\author[2,+]{R. M. Szmoski}
\author[3,+]{R. F. Pereira}
\author[4,+]{S. E. de Souza Pinto}
\affil[1]{Institute for Complex Systems and Mathematical Biology, SUPA, University of Aberdeen, Aberdeen, United Kingdom}
\affil[2]{Department of Physics, Federal University of Technology -  Paran\'a, 84016-210, Ponta Grossa, Paran\'a, Brazil}
\affil[3]{Department of Mathematics, Federal University of Technology - Paran\'a, 84016-210, Ponta Grossa, Paran\'a, Brazil}
\affil[4]{Departamento de F\'isica, Universidade Estadual de Ponta Grossa, 84030-900, Paran\'a, Brazil}

\affil[*]{murilo.baptista@abdn.ac.uk}

\affil[+]{these authors contributed equally to this work}

\keywords{Chaos, information, synchronisation, multiplex networks}

\begin{abstract}
The understanding of the relationship between topology and behaviour in interconnected networks would allow to characterise and predict behaviour in many real complex networks since both are usually not simultaneously known. Most previous studies have focused on the relationship between topology and synchronisation. In this work, we provide analytical formulas that shows how topology drives complex behaviour: chaos, information, and weak or strong synchronisation; in multiplex networks with constant Jacobian. We also study this relationship 
numerically in multiplex networks of Hindmarsh-Rose neurons. Whereas behaviour in the analytically tractable network is a direct but not trivial consequence of the spectra of eigenvalues of the Laplacian matrix,
where behaviour may strongly depend on the break of symmetry in the topology of interconnections, in Hindmarsh-Rose neural networks the nonlinear nature of the chemical synapses breaks the
elegant mathematical connection between the spectra of eigenvalues of the Laplacian matrix and the behaviour of the network, creating networks whose behaviour strongly depends on the nature
(chemical or electrical) of the inter synapses.
\end{abstract}
\begin{document}

\flushbottom
\maketitle
%
%
\thispagestyle{empty}

\section*{Introduction}

Complex networks \cite{Strogatz,Barabasi,ErdosRenyi} serve as a model for a broad range of phenomena. Brain \cite{Olaf1,sporns_PLOSBIOLOGY2008}, social interactions \cite{Carron}, and linguistics \cite{Dictionary} are all examples of systems represented by complex networks. In general, networks are useful models for studying systems that have a spatial extension. For instance,  insect populations whose interaction between them produces the extinction of one of them \cite{Rodrigo1}, the interaction between proteins \cite{Protein} and the interaction between gears \cite{Silvio}. These networks can be represented by a multiplex network of coupled complex subnetworks \cite{arenasprl,hernandez,arenas_PRE2013,multiplex1,multiplex2,multiplex3,multiplex4,multiplex5}. 

In the case of the brain \cite{sporns_PLOSBIOLOGY2008}, interconnections between complex subnetworks are typically made by chemical synapses while intraconnections can be formed by both chemical and electric synapses \cite{gallos}. For brain research \cite{gallos,chris} and brain-based cryptography \cite{criptoRomeu}, the interest is to understand the inter and intracouplings such that the units in the complex networks are sufficiently independent (unsynchronous) to achieve independent computations. However, the networks must be sufficiently connected (synchronous) such that information is exchanged between subnetworks and integrated into coherent patterns \cite{meunier2010}.

The academic community has dedicated much attention to elucidate the interplay between topology and behaviour in multiplex networks. In particular, the action of the inter and intracoupling strengths in the synchronisability of optimally evolved multiplex network graphs \cite{sarika}, and in the synchronisation of multiplex networks of dynamical oscillators \cite{arenasprl, zhao, asheghan2013, lu2010, guan2008} or neurons \cite{baptista_PRE2010,gallos, boccaletti, kurths2011}. Authors have shown an intricate interplay between different aspects of the network topology with weak or strong (not full) synchronisation, which was shown to be dependent on  the ratio between interlinks with all the links in networks of phase oscillators \cite{guan2008}, on the number of interlinks in networks of R\"ossler oscillators \cite{zhao} and neural networks \cite{kurths2011}, and on the ratio between inter and intra links in networks of heterogeneous maps \cite{lu2010}. Synchronisation was also shown to depend exclusively or complementarily on the electric or chemical couplings in two coupled neurons \cite{pfeuty2005} and in neural networks \cite{baptista_PRE2010,kurths2011,chris,arxiv-chris}. In particular, in the work of Ref. \cite{baptista_PRE2010}, it was shown semi-analytically that the stability of the complete synchronous manifold depends on the Laplacian matrix of the electric synapses, the degree of chemical synapses, and the type of chemical synapses (inhibitory or excitatory). The relationship between topology and the diffusive behaviour in multiplex networks composed by two coupled complex networks of ODEs with constant Jacobian was made clear in Refs. \cite{arenasprl}. 

In this work, we  
elucidate the interplay among the topological aspects previously described to be relevant in the study of synchronisation (i.e., the eigenvalues of the Laplacian, the ratio $\alpha$ between inter degree and the number of nodes of the subnetworks, and the inter and intra coupling strengths) and complex behaviour in multiplex networks of two undirected coupled equal complex networks. We will show analytically how topology drives and is related not only to weak or strong forms of synchronisation, but also to other complex forms of behaviour: chaos and information transmission. {\it Thus, providing an innovative set of mathematical tools to study how  complexity behaviour emerges in multiplex networks.} This achievement was possible because we were able to analytically calculate, for the first time, one of the most challenging quantities in nonlinear systems, the complete spectrum of Lyapunov Exponents for a class of multiplex networks with constant Jacobian. This intricate relationship was also studied numerically in multiplex neural networks.   

Our results show that in fact the ratio $\alpha$ is the determinant factor for the complex behaviour of the network, which also explain why the ratio between inter and intra or the number of interlinks  has been previously seem to drive synchronisation \cite{guan2008},\cite{zhao},\cite{lu2010},\cite{kurths2011}. We also show that synchronisation and information, whose quantifiers depend on the spectral gap of the Laplacian, will depend exclusively or complementarily on the inter and intra coupling strengths as observed in \cite{pfeuty2005,kurths2011}, and demonstrated in 
\cite{baptista_PRE2010}. For networks with constant Jacobian, synchronisation and information will depend exclusively on either the intra or the intercoupling strengths, 
if the two networks have symmetric interconnections, and will depend complementarily on  both intra and interconnections, if the  
two networks have asymmetric interconnections. For the multiplex neural networks, we find that intra and inter couplings will complementarily cooperate to complex behaviour if the two neural complex networks are coupled by inter chemical and excitatory synapses. If intercouplings are of the inhibitory nature, behaviour will mainly depend on the intracoupling. Therefore, it is the excitatory chemical synapses that promote integration between intra (local) and inter (global) synapses in neural networks. On the other hand, in the networks with constant Jacobian, integration between inter and intra comes about by the break of symmetry caused by the asymmetric configuration. Moreover, for this configuration, a bottle-neck effect appears for an appropriately rescaled intercoupling strength. In this case, an increase in the synchronisation level of the network leads to an increase in the capacity of the network to exchange information.

\section*{Methods}

\begin{figure}[hbt!]
\begin{center}
 \includegraphics[width=8cm,height=8cm]{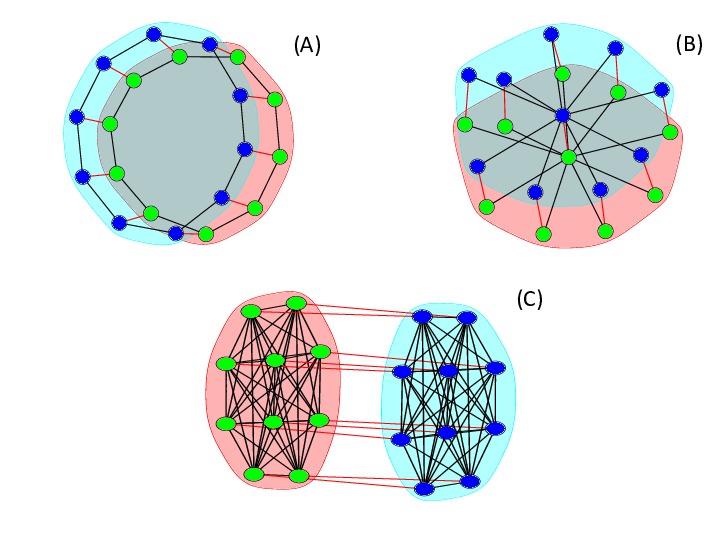}
 \end{center}
 \caption{[Colour online] Examples of symmetric network  topologies with $N=10$ and $\ell_{12}$=10 considered in this work. Subnetworks have a ring topology in (A), a star topology in (B), and an all-to-all topology in (C). Black lines represent intra links, and  Gray (red online) lines represent inter links.}
 \label{fig0}
\end{figure}

Each complex network connects with each other in two ways, by a symmetric or an asymmetric interlink configuration. For the symmetric case, each node in a subnetwork can have at most one connection with a corresponding node in the other equal subnetwork (See Fig. \ref{fig0}). The general asymmetric configuration presents nodes in one network that can randomly connect to other nodes in the other network. The considered network configurations are models of extended space-time chaotic systems \cite{Ahlers-Pikovsky,Pikovsky,Cencini,Tessone} or chemical chaos \cite{kiss1,kiss2}. It is also a model for two types of structures found in real neural networks \cite{deLange_FCN2014}. The one with stronger community structure (small first eigenvalue of Laplacian matrix, or strong intracouplings), and the one with a high level of bipartiteness, i.e., two similar complex networks strongly connected by intercouplings (larger last eigenvalue of the Laplacian matrix, or  strong intercoupling).

We consider two types of dynamics for the nodes of the network. The shift map (see Sec. "Extension to continuous networks" for networks with continuous-time descriptions), forming a discrete network of diffusively connected nodes, and the Hindmarsh-Rose (HR) neuron \cite{hr}, connected with inter chemical and intra electrically synapses.

Let $X$ represents the state variables of a network with $N=2N_1$ nodes formed by two equal coupled complex networks composed each of $N_1$ nodes that are coupled by $\ell_{12}$ "long-range" inter-connections.  
The dynamical description of the nodes is given by either the discrete-time function $F(x_n^{(i)})=2x_n^{(i)} (\mbox{mod 1}$) or the continuous-time function 
$\mathbf{f}(\mathbf{x}_i)$, representing the Hindmarsh-Rose neuron model.  

The discrete network of shift maps is described by 
{\small 
\begin{equation}
x_{n+1}^{(i)}=2x_n^{(i)} - {\varepsilon} \sum_{j=1}^{N}{G}_{ij}x_n^{(j)} - 
\gamma \alpha \sum_{j=1}^{N} {L}_{ij}x_n^{(j)}\,\mbox{(mod 1)},  
\label{network_discrete}
\end{equation}}where $\alpha=\frac{\ell_{12}}{N_1}$ represents an effective inter degree of the network. The network can 
be written in a matricial form by $\mathbf{x}_{n+1} = 2\mathbf{x}_n - [\varepsilon \mathbf{G} + \gamma \alpha \mathbf{L}]\mathbf{x}_n\,\mbox{(mod 1)}$, where 
$\mathbf{x}_n=[x_n^{(1)}\, x_n^{(2)} \ldots x_n^{(N)}]^{T}$,
$\mathbf{G} = \left( 
\begin{array}{cc}
\mathbf{A}  &  0  \\
0  & \mathbf{A}  
\end{array}
\right) $
and
 $\mathbf{L} = \left( 
\begin{array}{cc}
  \mathbf{D}_1 &  -\mathbf{B}   \\
 -\mathbf{B}^T   & \mathbf{D}_2  
\end{array}
\right) $ 
are Laplacian matrices and $T$ stands for the transpose. $\mathbf{G}$ represents the Laplacian of the two uncoupled complex networks and its intra links (the Laplacian matrix $\mathbf{A}$) and 
$\mathbf{L}$ represents the inter-couplings Laplacian matrix between the complex networks. $\mathbf{D}_1$ and $\mathbf{D}_2$ represent the identity degree of the adjacency matrices 
$\mathbf{B}$ and $\mathbf{B}^T$, respectivelly, representing the inter couplings. 
Their components are defined as $(\mathbf{D}_1)_{ii}=\sum_j {B}_{ij}$ and $(\mathbf{D}_2)_{ii}=\sum_j {B}^T_{ij}$, with null off diagonal terms. 
It can be written in an even more compact form by 
\begin{equation}
\mathbf{x}_{n+1} = 2\mathbf{x}_n - \mathbf{M}\mathbf{x}_n \mbox{(mod 1)}, 
\label{compact-discrete}
\end{equation}
\noindent
where 
$
\mathbf{M} = \left( 
\begin{array}{cc}
\varepsilon \mathbf{A} + \gamma \alpha \mathbf{D}_1 & -\gamma \alpha \mathbf{B} \\
-\gamma \alpha \mathbf{B}^T  & \varepsilon \mathbf{A} + \gamma \alpha \mathbf{D}_2 
\end{array}
\right)
$.

The network HR neurons represented by the coupling in the first coordinate is described by 
\begin{equation}
{\dot{x}^{(i)}_1}=f_1(\mathbf{x}^{(i)}) - {\varepsilon} \sum_{j=1}^{N}G_{ij} x^{(j)}_1 - 
\gamma (x^{(i)}_1-V_{syn}) \sum_{j=1}^{N} C_{ij}S(x^{(j)}_1), 
\label{network_continuous}
\end{equation}
\noindent
where $f_1$ represents the first component of the HR vector flow dynamics, $\mathbf{x}^{(i)}$ is a vector with components $(x^{(i)}_1,x^{(i)}_2,x^{(i)}_3)$ representing the variables of neuron $i$, $\mathbf{G}$ is the Laplacian for the intra electrical couplings, and $\mathbf{C}$ (with components ${C}_{ij}$) is an adjacency matrix representing the inter chemical couplings. The chemical synapses function $S$ is modelled by the sigmoidal function
$
S(x_1)=\displaystyle\frac{1}{1+e^{-\lambda(x_1 -\Theta_{syn})}},
$
\noindent
with $\Theta_{syn}=-0.25$, $\lambda=10$ and $V_{syn}=2.0$ for
excitatory and $V_{syn}=-2.0$ for inhibitory.  

In the brain, short-range connections among neurons happen by electric synapses, due to the potential difference of two neighbouring neuron body cells. In this work, the intra electrical synapses are mimicking this local interaction. Long-range connections are done by the chemical synapses, the inter connections in this work. However, to compare results between the HR networks and the discrete networks, the two subnetworks of HR neurons will have equal topologies, a configuration unlikely to be found in the brain. but that can however be interpreted as paradigmatic models of small brain circuits.  

As a measure of chaos, we consider the sum of the positive Lyapunov exponents of the network, denoted by $H_{KS}$. 
As a measure of the ability of the network to exchange information, we consider an upper bound for the Mutual Information Rate (MIR) between any two nodes in the network:
\begin{equation}
I_C = \lambda_1 - \lambda_2
\label{IC}
\end{equation}   
\noindent
in which $\lambda_1$ and $\lambda_2$ represent the two largest positive Lyapunov exponents of the network. We assume that these two largest Lyapunov exponents are approximations for the two largest expansion rates (or finite-time finite-resolution Lyapunov exponents) calculated in a bi-dimensional space \cite{baptista_PLOSONE2012} composed by any two nodes of the network. Equation (\ref{IC}) is constructed under the hypothesis that given two time-series, $x_1(t)$ and $x_2(t)$, an observer is not able to have a infinite resolution measurement of a trajectory point, but can only specify the location of a $x_1 \times x_2$ point within a 
cell belonging to an order-$T$ Markov partition, and thus the correlation of points decay to approximately zero after $T$ iterations.
For dynamical networks such as the ones we are working with, measurements can be done with higher resolution and it is typical to expect that the expansion rates on any 2D subspace formed by the state variables of two nodes are very good approximations of the 2 largest Lyapunov exponents of the network. 
Such a choice implies that $I_C$ in Eq. (\ref{IC}) is an invariant of the network and it represents the maximal rate of mutual information that can be realised when measurements are made in {\it any} two nodes of the network, and no time-delay reconstruction is performed. Details about the equivalence between Lyapunov exponents and expansion rates can be seen in Refs. \cite{baptista_PLOSONE2012},  and an explicitly numerical comparison can be seen in Ref. \cite{baptista-arxiv}. An extension of Eq. (\ref{IC}) to measure upper bounds of MIR in larger subspaces of a network (composed by group of nodes or multivariable subspaces) can also be seen in Ref. \cite{baptista_PLOSONE2012}.

Synchronisation is detected by various approaches. Linear stability of the synchronous manifold for complete synchronisation in the discrete network will be calculated analytically. For both types of networks, the level of weak synchronisation will be estimated by the value of $H_{KS}$, since the higher $H_{KS}$ is (and the larger with respect to $I_C$), the less synchronous nodes in the network are. Notice also that if $H_{KS}=I_C$, the network is generalised synchronous and possesses only one positive Lyapunov exponent.   For the network of Hindmarsh-Rose neurons, we measure synchronisation by calculating the order parameter $r$ and the local order parameter $r_{link}$ as introduced in  Ref. \cite{jesus_PRL2007}, the order parameter calculated considering the phase difference between all pair of nodes in the continuous network, as an estimation for the synchrony level of the network. The phase $\phi_i$ of a node $i$ is calculated using the equation for its derivative $\dot{\phi} = \frac{\dot{x}_2 x_1 - \dot{x}_1 x_2}{x_1^2+x_2^2}$ derived in 
Refs. \cite{Pereira_PRE2007} and \cite{Pereira_PLA2007}.

\section*{Results}

\subsection*{Shift map networks}

To calculate the Lyapunov exponents of the discrete network (see Sec. "Extension to continuous networks" for an extension to continuous networks), 
we recall that since the map produces a constant Jacobian ($[2\mathbb{I}  + \mathbf{M}]$) the Lyapunov spectra of the synchronisation manifold described by $x_n=x^{(i)}_n = {x}^{(j)}_n$ is equal to the spectra of Lyapunov exponents of the network (where typically $x^{(i)}_n \neq {x}^{(j)}_n$). In addition, the Lyapunov exponents of the synchronisation manifold are simply the Lyapunov exponents of the Master Stability Function (MSF) \cite{MSF}, the equations that describe the variational equations of Eq. (\ref{network_discrete}) linearly expanded around the synchronisation manifold (assuming $x_{n}^{(i)}=x_n+\xi_n^{(i)}$) and diagonalised, producing $N$ equations in the $m$ eigenmodes: 
\begin{equation}
{\delta}_{n+1}^{m}=[2-\mu_{m}]\delta_n^{m},  
\label{MSF}
\end{equation}
\noindent
where $\mu_m$ represents the eigenvalues of $\mathbf{M}$ ordered by magnitude, i.e., $\mu_0=0 \leq \mu_1 \leq \mu_2,\ldots,\leq \mu_{N-1}$. 
The ordered Lyapunov exponents are given by the logarithm of the absolute value of the derivative of the MSF in  (\ref{MSF}), which leads to
\begin{equation}
{\lambda}_{m+1} = \log{\left| 2 - \mu_{m} \right|}, 
\label{Lyapunov_exponents}
\end{equation} 

In this work, we consider two network configurations. Firstly, the {\it symmetric} configuration, when the two networks are connected by $\ell_{12}$ undirected interlinks, and each node in a network connects to at most one corresponding node in the other subnetwork. Secondly, the {\it asymmetric} configuration, when the two networks are connected by only one 
undirected random interlink. So, $\mathbf{D}_1=\mathbf{D}_2$. 

For the {\it symmetric} configuration \cite{hernandez} (see also Ref. \cite{baptista_PRE2010}), we have that  
\begin{equation}
\begin{array}{ccc}
\tilde{\mu}_{2i} & = & \epsilon \omega_{i} \\
\tilde{\mu}_{2i+1} & = & \epsilon \omega_{i} + 2 \gamma \alpha 
\label{eigenvalues-symmetric}
\end{array}
\end{equation}
\noindent
where $\tilde{\mu}_{2i}$ are the unordered eigenvalues of $\mathbf{M}$ ($i=0,1,2,\ldots,N_1-1$) and $\omega_i$ represents the ordered set of eigenvalues of the Matrix $\mathbf{A}$ (such that $\omega_{i+1} \geq  \omega_{i}$, and $\omega_0=0$), whose unordered spectra is given by  
$\tilde{\omega}_{i} = 2\left[1 - \cos{\left( \frac{2\pi i}{N_1} \right)}\right]$ for a closed ring topology,
or $\omega_1=0$, $\omega_i=1$ (for $i=1,\ldots,N_1-2$), and $\omega_{N-1}=N_1$ for a star topology, 
and $\omega_1=0$, $\omega_k=N_1$, for all-to-all topology.   The inter degree $\alpha$ represents the 
effective connection that every node in one subnetwork will have with the other. If $2\gamma\alpha < \epsilon \omega_1$, then $\mu_1= 2\gamma\alpha$, otherwise  $\mu_1= \epsilon \omega_1$.
Complete synchronisation of the shift map network is linearly stable if $|2-\mu_1|<1$, however notice that 
our study considers coupling ranges outside of the complete stability region. The second largest eigenvalue, $\mu_1$, and therefore $I_C$ (and the stability of the synchronous manifold)  will only depend on the inter connections if  
\begin{equation}
\gamma < \frac{\epsilon \omega_1}{2\alpha}, 
\label{condition_gamma}
\end{equation}
\noindent
and these quantities will only depend on the intra connections if this inequality is not satisfied.  

It is fundamental to mentioning that the eigenvalues obtained in Eq. (\ref{eigenvalues-symmetric}) using the expansion in \cite{hernandez} provide values that are exact in the topologies considered in this work (demonstration to appear elsewhere). Consequently,  the Lyapunov exponents calculated by Eq. (\ref{Lyapunov_exponents}) are also exact. 

For the {\it symmetric} configuration, 
\noindent
if inequality (\ref{condition_gamma}) is satisfied, $\lambda_2=\log{\left | 2(1 - \gamma \frac{\ell_{12}}{N_1}) \right |}$, or  
$\lambda_2=\log{\left | (2 - \epsilon  \omega_1 ) \right |}$, otherwise. Since $\lambda_1=\log{(2)}$, then the upper bound for the MIR 
exchanged between any two nodes in this network, assuming $\lambda_2>0$, is given by 
\begin{eqnarray}
I_C & = & -\log{\left(1- \gamma \frac{\ell_{12}}{N_1} \right)} \mbox{, if inequality (\ref{condition_gamma}) satisfied}  \\
I_C & = & -\log{\left(1- \epsilon \frac{\omega_1}{2} \right)} \mbox{, otherwise.} 
\label{I_C_maps}
\end{eqnarray}
\noindent 
Therefore, the upper bound for the MIR will either depend on $\gamma$ or on $\epsilon$. If $\lambda_2\leq 0$, then $I_C=\lambda_1=\log{(2)}$.

For the {\it asymmetric} configuration 
\cite{hernandez}, we have that  
\begin{equation}
\begin{array}{ccc}
\tilde{\mu}_{0} & = & \epsilon \omega_{0} \\
\tilde{\mu}_{1} & = & 2\gamma \alpha \\
\tilde{\mu}_{2i} & = & \epsilon \omega_{i} + \gamma \alpha \\
\tilde{\mu}_{2i+1} & = & \epsilon \omega_{i} + \gamma \alpha \\
\end{array}
\label{bottle-neck}
\end{equation}
\noindent 
for $i=1,2,\ldots,N_1-1$. If $ \tilde{\mu}_1 <  \tilde{\mu}_{2}$, then $\mu_1=  \tilde{\mu}_1 $ and 
$\mu_2= \tilde{\mu}_2$, otherwise 
$\mu_1= \tilde{\mu}_2$ and $\mu_2= \tilde{\mu}_1$. Complete synchronisation is linearly stable if $|2-\mu_1|<1$. 
If 
\begin{equation}
\gamma < \frac{\epsilon \omega_1}{\alpha}, 
\label{condition_gamma1}
\end{equation}
\noindent
the second largest eigenvalue and, therefore, $I_C$ (and the stability of the synchronous manifold) will only depend on the interconnection. If this inequality is not satisfied, these quantities will depend mutually on both types of connections. Since $\alpha$  
always appears in the second largest eigenvalue, the smallest its value the largest will be $I_C$.  
Our analytical results are valid for all asymmetric configurations considered in Ref. \cite{hernandez}, however in this paper we focus on the   "bottleneck" configuration, where there is only one random interlink.

\begin{figure}[hbt!]
\begin{center}
 \includegraphics[width=8.5cm,height=9.7cm]{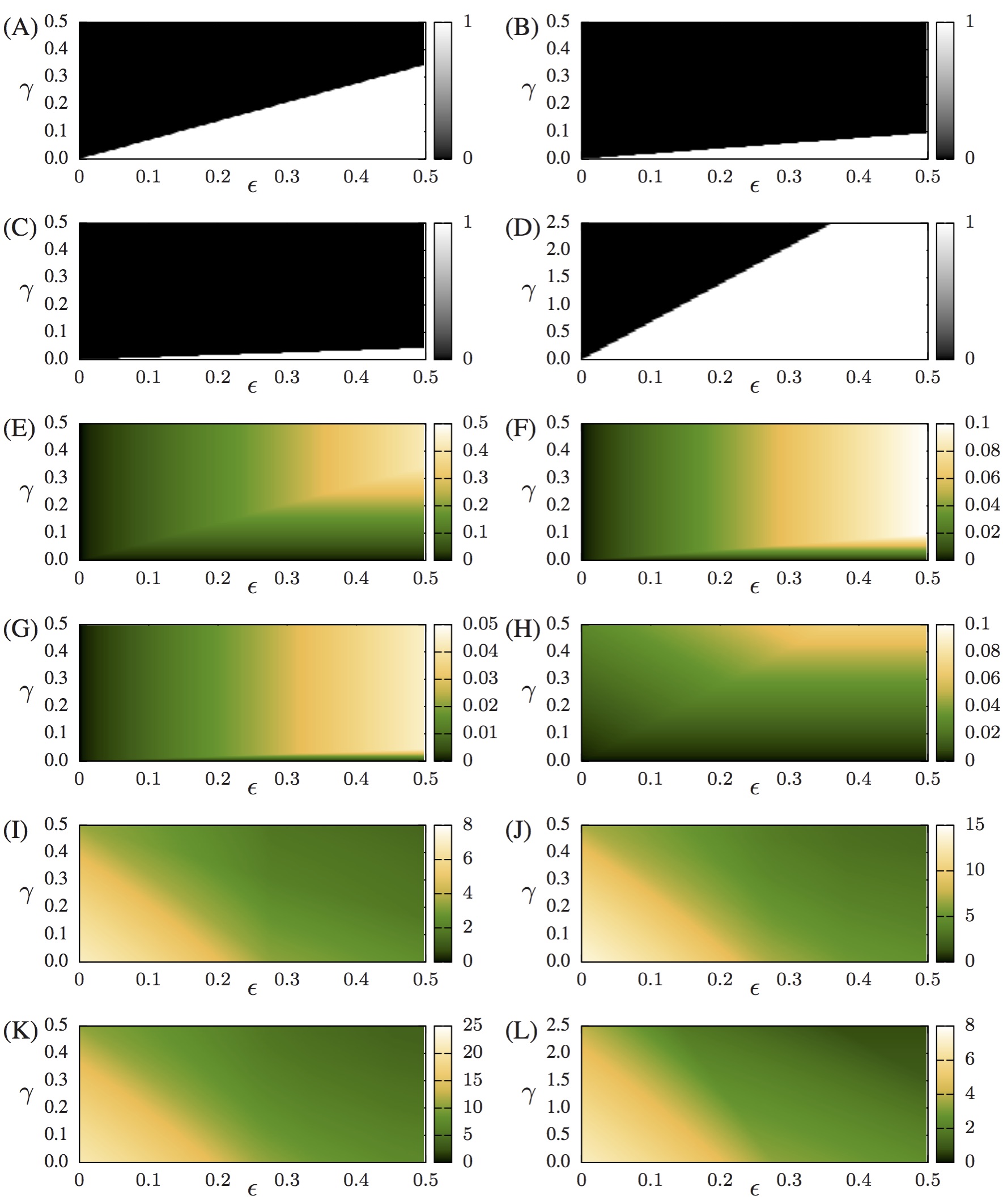}
 \end{center}
 \caption{[Colour online] Results for networks of shift maps, with two coupled rings. (A-C) White (black) region indicates values of $\epsilon$ and $\gamma$ for which inequality (\ref{condition_gamma}) is satisfied (not satisfied). (D) Colour code same as in (A-C), but based on inequality (\ref{condition_gamma1}). (E-H) color code shows the value of $I_C$. (I-L) Sum of positive Lyapunov exponents.  $N$=10 and $\ell_{12}$=5 in (A), (E) and (I), $N$=20 and $\ell_{12}$=10 in (B), (F), 
(J), $N$=30 and $\ell_{12}$=15 in (C), (G), and (K),  $N$=10 and  $\ell_{12}$=1 in (D), (H), and (L). In (L), 
the maximal value of $\gamma$ equal to 2.5 was chosen to allow that 
the range of values for the quantity $\gamma\alpha$ is the same in figures (I)-(L).}
 \label{fig1-ring}
\end{figure}

For the asymmetric {\it bottle neck} configuration, 
\noindent
if inequality (\ref{condition_gamma1}) is satisfied, $\lambda_2=\log{\left | 2(1 - \gamma \frac{\ell_{12}}{N_1}) \right |}$, or  
$\lambda_2=\log{\left | (2 - \epsilon  \omega_1 -\gamma\frac{\ell_{12}}{N_1}) \right |}$, otherwise. Since $\lambda_1=\log{(2)}$, then the upper bound for the MIR 
exchanged between any two nodes in this network, assuming $\lambda_2>0$,  is given by 
\begin{eqnarray}
I_C & = & -\log{\left(1- \gamma \frac{\ell_{12}}{N_1} \right)} \mbox{, if inequality (\ref{condition_gamma1}) satisfied}  \\
I_C & = & -\log{\left(1- \epsilon \frac{\omega_1}{2} - \gamma \frac{\ell_{12}}{2N_1} \right)} \mbox{, otherwise.} 
\label{I_C_maps2}
\end{eqnarray}
\noindent 
Therefore, the upper bound for the MIR will either depend on $\gamma$, if   inequality (\ref{condition_gamma1}) is satisfied, or on both couplings if this inequality is not satisfied.  If $\lambda_2\leq 0$, then $I_C=\lambda_1=\log{(2)}$.

 Figures \ref{fig1-ring}(A-D) are parameter spaces ($\epsilon$ $\times$ $\gamma$) showing whether inequality (\ref{condition_gamma}) (A-C) or inequality (\ref{condition_gamma1}) (D) are satisfied (white) or not (black). Figures \ref{fig1-ring}(E-H) show the value of $I_C$. In Figs. \ref{fig1-ring}(E-G) we show results for the symmetric configuration.  $I_C$ will only depend on the intercoupling $\gamma$ if inequality  (\ref{condition_gamma}) is satisfied, and will only depend on the intracoupling $\epsilon$ if this inequality is not satisfied. In Figs. \ref{fig1-ring}(H), for the bottleneck configuration, $I_C$ will only 
depend on the inter coupling if this inequality is satisfied, but will depend on  both inter and intra couplings if this inequality is not satisfied.   
The sum of Lyapunov exponents is given by $H_{KS}=P \log{(2)}+\sum_{m=1}^P \log{(|1-\mu_m/2|)}$, where $P$ represents the number of positive Lyapunov exponents of the network. From this equation, it becomes clear that if $N_1$ is increased and the topology considered makes $\omega_i$ to increase proportional to $N_1$, but the ratio $\alpha$ is maintained (meaning that inter connections grow only proportional to $N_1$), then the term $\epsilon \omega_i$ becomes predominant in $H_{KS}$, and as a consequence, chaos in the network becomes more dependent on $\epsilon$ than on $\gamma$. To illustrate this argument, let us consider the symmetric configuration and assume that $\epsilon$ and $\gamma$ are sufficiently small such that all Lyapunov exponents are positive. Then, the summation to calculate $H_{KS}$ has $N$ terms and $H_{KS}=N\log{(2)}-\sum_{i=0}^{N_1-1}(\epsilon \omega_i + \gamma \alpha)$. Thus, the term with $\epsilon$ dominates for larger $N_1$. This becomes even more evident, if the topology is an all-to-all: $H_{KS}=N\log{(2)-(N_1-1)N\epsilon}-\gamma \ell_{12}$. The predominance of the intra coupling can be seem in all panels of Fig. \ref{fig1-ring}(I-L), for a network of two coupled ring networks. 
Similar results to other network configurations can be seen in Supplementary Material. 

\subsubsection*{Extension to continuous networks}

These results can be extended to linear networks of ODEs. As an example, consider a continuous network of 1D coupled linear ODEs described by 
$ \dot{\vec{x}} =  [\alpha \mathbb{I}+ \mathbf{M}] \vec{x}$. Then, the Lyapunov exponents of this system are equal to the Lyapunov exponents of the synchronisation manifold and its transversal modes, and therefore are equal to $\lambda_{m+1} = \alpha - \mu_{m}$.

\subsection*{Hindmarsh-Rose networks}

The Lyapunov Exponents of the HR neural networks are calculated numerically.

\begin{figure}[hbt!]
\begin{center}
 \includegraphics[width=8.5cm,height=9.7cm]{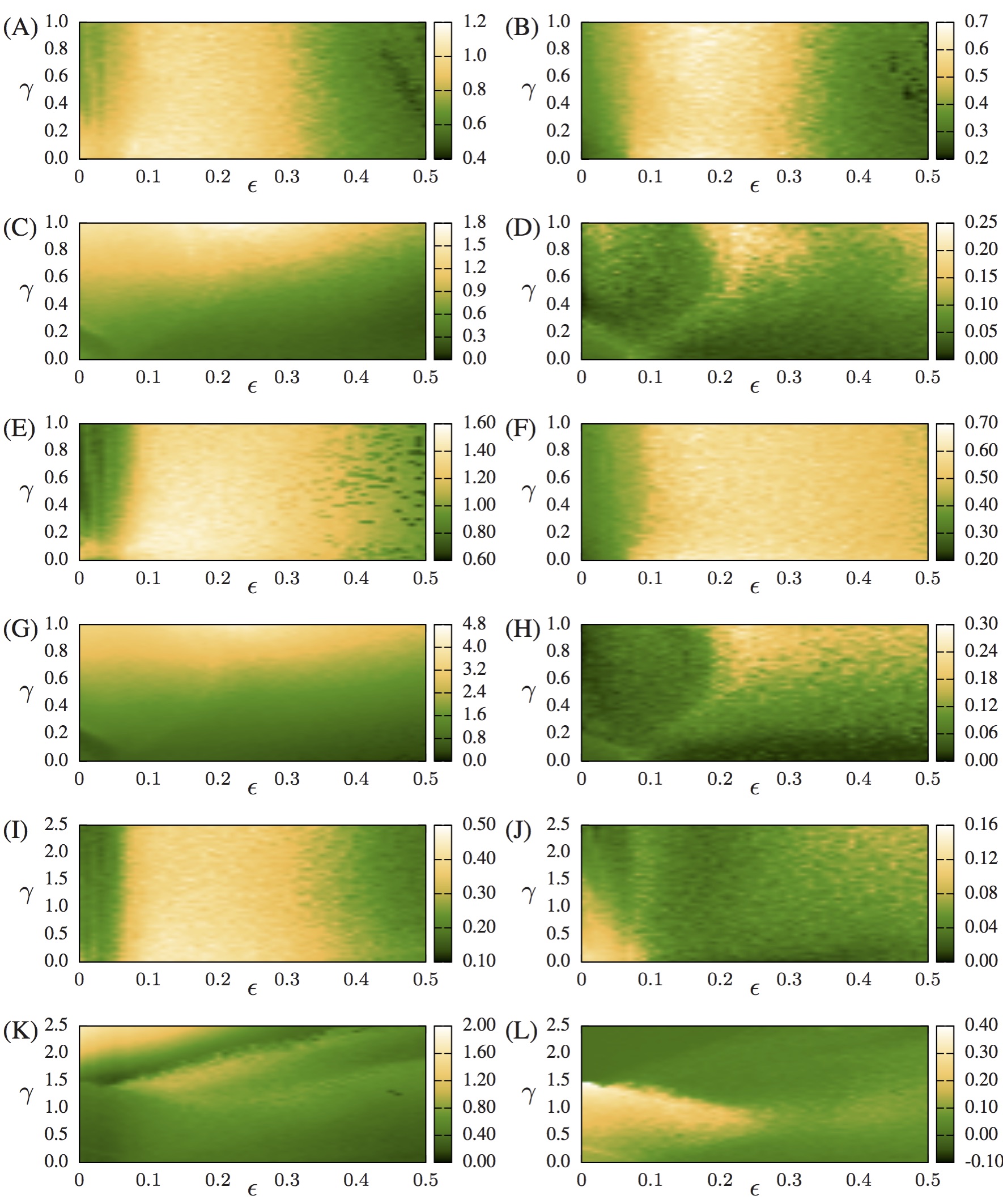}
 \end{center}
 \caption{$N$=10 and $\ell_{12}$=5 for (A-D), $N$=20 and $\ell_{12}$=10 for (E-H), and  $N$=10 and $\ell_{12}$=1 for (I-L). 
 Sum of positive Lyapunov exponents shown in left column and 
 $I_C$ shown in right column for 
 two coupled rings of Hindmarsh-Rose neurons with inter inhibitory coupling (in (A), (B), (E), (F), (I), and (J)) and 
inter excitatory coupling (in (C), (D), (G), (H), (K), (L)).}
 \label{fig2}
\end{figure}

For symmetric HR neural networks with inhibitory inter connections, $H_{KS}$ is mostly dependent only on the electrical intra coupling, as can be seen from Figs. \ref{fig2}(A),(E),(I) (for coupled ring complex networks), and the results shown in Supplementary Material, for other networks. The quantity $I_C$ is also mostly dependent on the electrical intra coupling in asymmetric  configurations with inter inhibitory synapses (see Fig. \ref{fig2}(B) and Fig. \ref{fig2}(F)), 
but for the asymmetric and inhibitory configuration (Fig. \ref{fig2}(J)), $I_C$ values depend mutually on both inter and intra couplings. 
Therefore, in most of the cases studied, neural networks formed by complex networks connected with inhibitory connections will have a behaviour
($H_{KS}$ and $I_C$) that mainly depends on the intra electric coupling. 
If inter connections are excitatory, both $H_{KS}$ and $I_S$ are a non-trivial function of the inter and intra coupling, as it can be seen in Figs.  \ref{fig2}(C-D),(G-H), (K)-(L).  The inter degree $\alpha$ is also determinant for the similar behaviours observed in symmetric neural networks (for both inhibitory and excitatory) of different sizes, as one can check by verifying how similar the parameter spaces of  Figs. \ref{fig2}(A-B) are with the ones in Figs. \ref{fig2}(E-F), or 
the parameter spaces of Figs. \ref{fig2}(C-D) and the ones in Figs. \ref{fig2}(G-H)).

To understand why  if 
different neural networks have equal inter-degree $\ell_{12}/N_1$, then they will have similar parameter spaces for  
$H_{KS}$ and $I_C$, we consider the conjecture of Ref. \cite{baptista_PLA2011} that shows that Lyapunov exponents and  Lyapunov Exponent of the synchronisation manifold (LESM) (defined by $\bf{x}^{(i)}=\bf{x}^{(j)}=\bf{x}^s$) are connected. Then, we remark that if each neuron in the network has the same inter-degree,  $k$, then $\ell_{12}/N_1=k$. This is a necessary condition in order to  obtain a Master Stability Function (MSF) of  the network as derived in \cite{baptista_PRE2010}. The linear stability of this network and the  $i$th  LESM  of this network will depend on a function $\Gamma =  - k \gamma S(x^s_{1}) - \gamma (x^s_{1} - V_{syn}) S^{\prime}(x^s_{1})(k - \mu^{\prime}_i) - \epsilon \omega_i$, where $\mu^{\prime}_i$ represents the eigenvalues of the Laplacian matrix $\bf{B}$. Inhibition or excitation contributes to the stability of the MSF and to the LESM through the term $\gamma (x^s_{1} - V_{syn}) S^{\prime}(x^s_{1})(k - \mu^{\prime}_i)$. 
If the coupling is inhibitory, all the terms in the function $\Gamma$ will be negative, and they all typically contribute to making the network more stable and to have smaller values of LESM. But both terms, $k \gamma S(x^s_{1})$ and $\gamma (x^s_{1} - V_{syn}) S^{\prime}(x^s_{1})(k - \mu^{\prime}_i)$, can be neglected, since $S$ is nonzero during a spike and $S^{\prime}$ is only nonzero at the moment of the beginning of a spike. Therefore, the stability of the synchronisation manifold, as well as the LEs and $I_C$ (using \cite{baptista_PLA2011}) will mainly depend on the value of the intra coupling $\epsilon$ (see also Fig. 5 in Ref. \cite{baptista_PRE2010}). If, however, the coupling is excitatory, we cannot neglect the term $\gamma (x^s_{1} - V_{syn}) S^{\prime}(x^s_{1})(k - \mu^{\prime}_i)$. If two networks with different sizes have the same $k$ for each neurone, then the eigenvalues of $\bf{B}$ for the larger network will be the same of the ones for the smaller network but appearing with multiplicity given by the dimension of the matrix. If the two different networks have the same topology, then some of the smallest eigenvalues of $\bf{A}$ for the larger network might be similar. These smallest eigenvalues contribute to making the term   $\epsilon \omega_i$  small, but with a magnitude  comparable to the magnitude of the term $\gamma (x^s_{1} - V_{syn}) S^{\prime}(x^s_{1})(k - \mu^{\prime}_i)$. Thus, if $k$ is made constant, larger networks might present similar parameter spaces for $H_{KS}$ and $I_C$.

\subsection*{The bottleneck effect}

In the bottleneck configuration, the inter-degree decreases to $1/N_{1}$.  This results in a value of $\gamma \alpha$ smaller when compared to this value for symmetric configurations. Consequently, given two networks, one symmetric and another asymmetric, both with the same $N_1$ and the same $\gamma \lambda_2$, the value of $\lambda_2$ for the asymmetric bottleneck configuration will be larger than $\lambda_2$ for the symmetric configuration, which leads to that $I_C$ for the asymmetric case is smaller than $I_C$ for the symmetric case. However, if we rescale $\gamma$ used in the 
asymmetric bottleneck configuration to keep the quantity $\gamma \alpha$ constant in all our simulations, the term $\epsilon \omega_1$ appearing in 
$\mu_1$ will  compensate $\lambda_2$ when inequality (\ref{condition_gamma1}) is satisfied, finally producing an asymmetric network that has a larger value of $I_C$ than the corresponding symmetric one. Regarding the neuronal networks, the bottleneck effect is evident as one compare  
Fig.  \ref{fig2}(L) (asymmetric) with Figs. \ref{fig2}(D) and  \ref{fig2}(F). No bottleneck effect was verified for inhibitory inter synapses. Concluding, a decrease in synchronisation can increase the capacity of the network to exchange information. 

\section*{Discussion}

A topic of research that has attracted great attention in multiplex networks was the search for a better understanding of how weak or strong synchronisation (not full) is linked to the various aspects of the network topology. Previous works have provided complementary, but not unified conclusions regarding this relationship. One of the difficulties into clarifying this matter is that the relationship between the spectrum of eigenvalues of the connecting Laplacian matrix and the synchronous behaviour of the network is poorly understood when the network is in a typical natural state and there is no full synchronisation.     
Our main contribution in this work was to understand this relationship when a multiplex network is out of full synchronisation. We went a step further and have also understood how relevant aspects of the network topology are related to chaos and information transmission.  {\it{Thus, providing an innovative set of mathematical tools to study how  complexity behaviour emerges in multiplex networks.} }


\section*{Acknowledgements}

MSB acknowledges the Engineering and Physical Sciences
Research Council grant Ref. EP/I032606/1. This work was also  partially supported by CNPq, CAPES, and Funda\c c\~ao Arauc\'aria.

\section*{Author contributions statement}

All authors have equally contributed to the conceptualisation of this work, numerical work and its analysis, and to the writing of the manuscript.  

\section*{Additional information}

\textbf{There are no competing financial interests}. 



\end{document}